\documentclass[final,5p,times,twocolumn]{elsarticle}

\usepackage[T1]{fontenc}
\usepackage[utf8]{inputenc}
\usepackage{amsmath,amssymb,amsfonts,bm,mathtools}
\usepackage{graphicx}
\usepackage{xcolor}
\usepackage{hyperref}
\hypersetup{colorlinks=true,linkcolor=blue,citecolor=blue,urlcolor=blue}
\biboptions{sort&compress}
\journal{Physics Letters B}

\begin{document}

\begin{frontmatter}

\title{Signatures of a de Sitter-core black hole in ringing, transmission and optical appearance}

\author[khazar,damghan,campina]{N. Heidari\corref{cor1}}
\ead{heidari.n@gmail.com}
\author[ufpb,campina,khazar]{A. A. Ara\'{u}jo Filho}
\ead{dilto@fisica.ufc.br}
\author[sao,herzen,khazar]{V. Vertogradov}
\ead{vdvertogradov@gmail.com}
\author[emu]{A. \"Ovg\"un}
\ead{ali.ovgun@emu.edu.tr}

\cortext[cor1]{Corresponding author.}

\address[khazar]{Center for Theoretical Physics, Khazar University, 41 Mehseti Street, Baku, AZ-1096, Azerbaijan}
\address[damghan]{School of Physics, Damghan University, Damghan, 3671641167, Iran}
\address[ufpb]{Departamento de F\'{i}sica, Universidade Federal da Para\'{i}ba, Caixa Postal 5008, 58051-970, Jo\~{a}o Pessoa, Para\'{i}ba, Brazil}
\address[sao]{SPB Branch of SAO RAS, 65 Pulkovskoe Rd, Saint Petersburg 196140, Russia}
\address[herzen]{Physics Department, Herzen State Pedagogical University of Russia, 48 Moika Emb., Saint Petersburg 191186, Russia}
\address[emu]{Physics Department, Eastern Mediterranean University, Famagusta, 99628 North Cyprus, via Mersin 10, T\"urkiye}
\address[campina]{Departamento de Física, Universidade Federal de Campina Grande Caixa Postal 10071, 58429-900 Campina Grande, Paraíba, Brazil.}


\begin{abstract}
We investigate a static, asymptotically flat black hole whose central region approaches a de Sitter vacuum.  The geometry is controlled by the ADM mass $M_0$ and a core scale $R$, and is generated by an exponentially decaying anisotropic source.  After clarifying the stress tensor and the horizon structure, we study scalar-field perturbations, greybody transmission, Hawking emission and the optical image produced by an optically thin infalling flow.  The horizon analysis shows that two horizons exist below the critical value $R/M_0\simeq0.7768$, where they merge into an extremal configuration.  Increasing the core scale lowers the peak of the scalar effective potential and shifts the quasinormal spectrum away from the Schwarzschild value, with the largest fractional deviations occurring for low multipoles and higher overtones within the WKB domain of validity.  The greybody bound indicates stronger filtering as the core becomes more extended, while the QNM--greybody correspondence gives a complementary estimate of the transmission probability in the eikonal regime.  The Hawking temperature decreases as the extremal configuration is approached, suppressing the emission rate and shifting its maximum to lower frequencies. {\color{black}For a simple optically thin infalling flow, ray tracing shows a small decrease in the apparent shadow scale and in the peak intensity as the core scale is increased.}
\end{abstract}


\end{frontmatter}

\section{Introduction}
\label{sec:introduction}

Gravitational collapse is one of the central predictions of general relativity.  In the Oppenheimer--Snyder--Datt model, a spherical dust cloud collapses through its Schwarzschild radius and forms a black hole whose curvature singularity is hidden behind an event horizon \cite{r1,r2}.  Whether singularity formation is unavoidable, or whether quantum-gravity or high-density matter effects can replace the singular core by a regular region, remains an important question in black-hole physics.  This issue is also connected with cosmic censorship, the nature of ultra-compact objects, and the possible observational imprints of near-horizon and near-core deviations from the Schwarzschild geometry \cite{r3}.

A widely studied possibility is that sufficiently dense matter undergoes a transition to a vacuum-like state with equation of state $p=-\rho$.  This idea goes back to the proposals of Gliner and Sakharov, where the central region of a compact object can be modelled by a de Sitter core rather than by a curvature singularity \cite{bib:Gliner,bib:Sakharov}.  In such models the spacetime interpolates between a de Sitter-like interior and an asymptotically Schwarzschild exterior.  Related scenarios include regular black holes with Minkowski cores, wormhole-like cores and other non-singular completions; see Refs.~\cite{bib:review1,bib:review2} for reviews and further references.

The phenomenology of compact objects with regular cores is particularly timely.  Gravitational-wave observations have opened a direct window onto black-hole ringdown, while horizon-scale interferometry by the Event Horizon Telescope has measured the emission region around M87$^*$ and Sgr A$^*$ \cite{r17,EventHorizonTelescope:2019dse,EventHorizonTelescope:2022wkp}.  Black-hole shadow observables have been extensively investigated in non-Kerr geometries, plasma environments, quantum-corrected backgrounds, and horizon-scale tests of gravity, providing a powerful phenomenological bridge between strong-field theory and imaging observations \cite{Atamurotov:2013sca,Atamurotov:2015nra,Vagnozzi:2022moj,Wang:2025fmz,Battista:2026nsx}. These observations make it useful to ask how a regular core changes quantities that are directly tied to wave propagation and null geodesics, such as quasinormal modes (QNMs), greybody factors, Hawking emission and shadow observables \cite{r20,r21,r22,r23,r24,r25,r26,r27,r28}.  Even when the deviations are small outside the horizon, the effective potential governing perturbations and photon propagation can retain measurable information about the interior scale.

In this Letter we analyze a black hole with a de Sitter core described by the exact static solution of Ref.~\cite{vertogradov2024exact}.  We first review the geometry and correct the stress-tensor interpretation in a form suited for the present applications.  We then derive the master equation for massless perturbations, compute QNMs with the third-order WKB method, study semi-analytic greybody bounds and their approximate correspondence with QNM data, and discuss the Hawking temperature and emission spectrum.  We finally examine the optical appearance generated by an optically thin, radially infalling emitting flow.  Throughout we use units $c=\hbar=k_B=1$ and write the Einstein equations in the normalization $G^\mu{}_{\nu}=T^\mu{}_{\nu}$, i.e. the factor $8\pi G$ is absorbed into the definition of the effective stress tensor.

\section{Brief review of exact regular black hole solutions with de
sitter cores}
\label{sec:geometry}

The spacetime considered in Ref.~\cite{vertogradov2024exact} is static and spherically symmetric,
\begin{equation}
  \mathrm{d}s^2=-f(r)\mathrm{d}t^2+\frac{\mathrm{d}r^2}{f(r)}+r^2\mathrm{d}\Omega^2,
  \label{eq:metric}
\end{equation}
with
\begin{equation}
  f(r)=1-\frac{2M_0}{r}\left[1-\frac{2}{R^2}\left(r^2+rR+\frac{R^2}{2}\right)e^{-2r/R}\right].
  \label{eq:lapse}
\end{equation}
Equivalently, $f(r)=1-2m(r)/r$, where
\begin{equation}
  m(r)=M_0\left[1-\left(\frac{2r^2}{R^2}+\frac{2r}{R}+1\right)e^{-2r/R}\right].
  \label{eq:massfunction}
\end{equation}
The ADM mass is $M_0$, since $m(r)\to M_0$ at infinity.  Near the center,
\begin{equation}
  f(r)=1-\frac{8M_0}{3R^3}r^2+O(r^3),
  \label{eq:core_expansion}
\end{equation}
which is the de Sitter form with an effective central density scale $\rho_0=8M_0/R^3$ in the above normalization.

For the metric gauge in Eq.~\eqref{eq:metric}, the effective matter source is anisotropic.  Directly from the Einstein equations one obtains
\begin{equation}
  \rho=\frac{2m'(r)}{r^2}=\frac{8M_0}{R^3}e^{-2r/R},
  \qquad
  p_r=-\rho,
  \label{eq:rho_pr}
\end{equation}
where $p_r$ is the radial pressure.  The transverse pressure is
\begin{equation}
  p_\perp=-\frac{m''(r)}{r}=\left(\frac{r}{R}-1\right)\frac{8M_0}{R^3}e^{-2r/R}.
  \label{eq:pt}
\end{equation}
Thus the de Sitter equation of state is recovered at the center, $p_r(0)=p_\perp(0)=-\rho_0$.  Eliminating $r$ in favor of $\rho$ gives the covariant transverse equation of state
\begin{equation}
  p_\perp(\rho)=\rho\left[-1+\frac{1}{2}\ln\left(\frac{\rho_0}{\rho}\right)\right],
  \qquad \rho_0=\frac{8M_0}{R^3}.
  \label{eq:eos_covariant}
\end{equation}
The explicit factor of $\rho$ is required on dimensional grounds.

The horizons are the positive roots of $f(r)=0$.  It is useful to introduce $x=r/M_0$ and $a=R/M_0$.  In the Schwarzschild limit $a\to0$, the outer horizon tends to $r_+=2M_0$, while an inner horizon moves toward the origin.  The two horizons merge at an extremal point determined by $f(r_e)=f'(r_e)=0$.  A direct numerical solution gives
\begin{equation}
  \frac{R_e}{M_0}\simeq0.776789,
  \qquad
  \frac{r_e}{M_0}\simeq1.314185.
  \label{eq:extremal_values}
\end{equation}
Therefore black-hole configurations exist for $R/M_0<R_e/M_0$ and disappear above this critical value.

The function
\begin{equation}
  A(r)=\frac{2}{R^2}\left(r^2+rR+\frac{R^2}{2}\right)e^{-2r/R}
  \label{eq:A_function}
\end{equation}
satisfies $5e^{-2}\le A\le1$ for $0\le r\le R$.  Hence any horizon lying inside the scale $R$ obeys $r_h\le 2M_0(1-5e^{-2})\simeq0.647M_0$.  This is a bound on a root inside the matter scale, not on the outer event horizon, which approaches the Schwarzschild value for $R/M_0\ll1$. {\color{black} The present solution differs from the commonly used Bardeen and Hayward metrics in the form of its mass profile.  In those models the regular core is usually introduced through rational functions of the radial coordinate, whereas here the source density decays exponentially, $\rho\propto e^{-2r/R}$.  The model is closer in spirit to Dymnikova-type geometries, although the precise density profile and mass function are different.  It is also distinct from Simpson--Visser black-bounce spacetimes, where regularity is produced by modifying the areal radius rather than by an exponentially localized anisotropic source. These differences are expected to affect the horizon structure, photon-sphere position, effective potential, quasinormal spectrum, greybody transmission and optical shadow scale.}

\begin{figure}[t]
  \centering
  \includegraphics[width=0.92\linewidth]{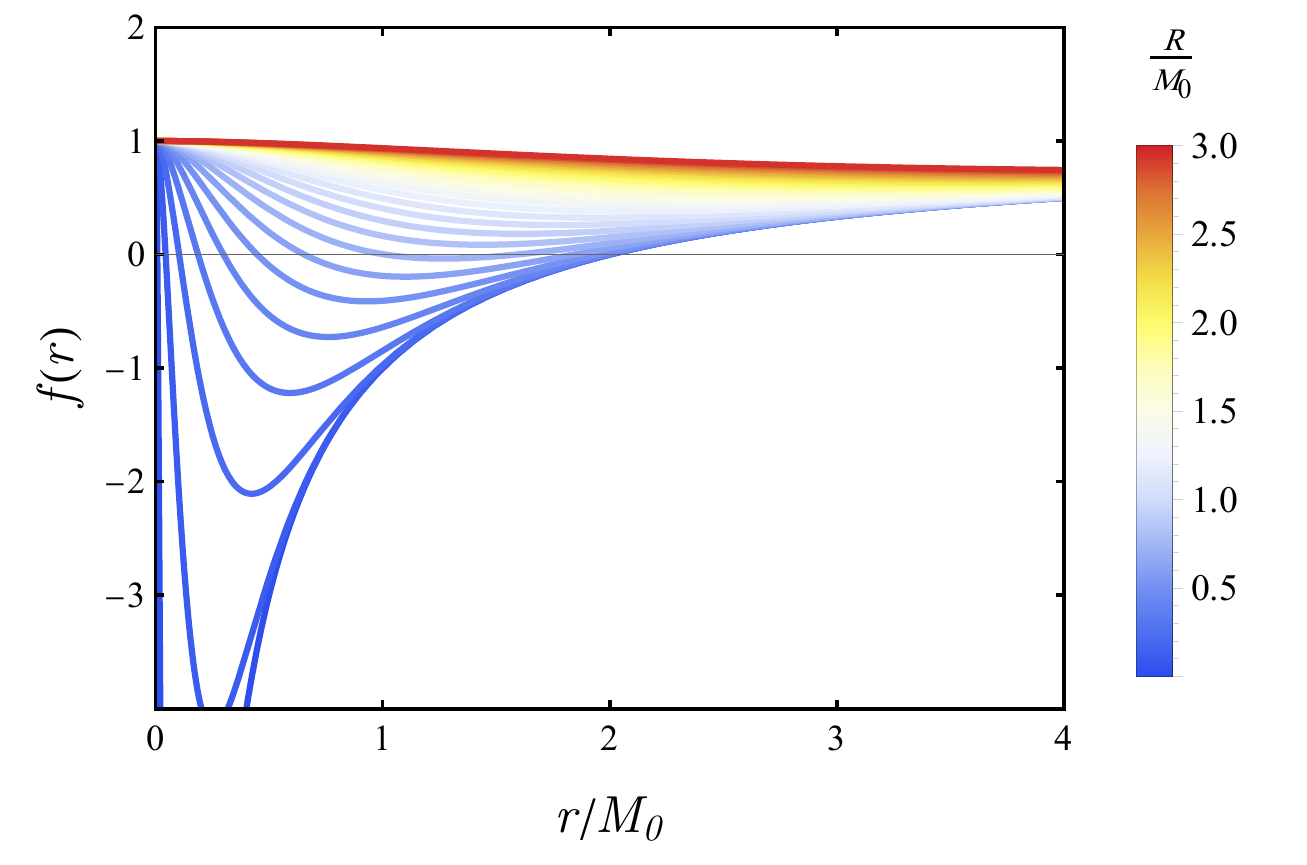}
  \caption{Lapse function for the Schwarzschild black hole and for the de Sitter-core black hole at representative values of $R/M_0$.  The two horizons merge at $R/M_0\simeq0.7768$.}
  \label{fig:lapse}
\end{figure}

We consider a massless perturbing field on the background \eqref{eq:metric}.  For the scalar case the Klein--Gordon equation is
\begin{equation}
  \frac{1}{\sqrt{-g}}\partial_\mu\left(\sqrt{-g}g^{\mu\nu}\partial_\nu\Psi\right)=0.
  \label{eq:kg}
\end{equation}
With
\begin{equation}
  \Psi_{\omega \ell m}(t,r,\theta,\phi)=\frac{\psi_{\omega\ell}(r)}{r}Y_{\ell m}(\theta,\phi)e^{-i\omega t}
  \label{eq:separation}
\end{equation}
and the tortoise coordinate $\mathrm{d}r_*/\mathrm{d}r=f^{-1}$, the radial equation assumes the Schr\"odinger form
\begin{equation}
  \left[\frac{\mathrm{d}^2}{\mathrm{d}r_*^2}+\omega^2-V_s(r)\right]\psi_{\omega\ell}=0.
  \label{eq:master}
\end{equation}
For spin $s=0,1$ test fields, and for a Regge--Wheeler-type axial gravitational potential, we use \cite{AraujoFilho:2024xhm,Chen:2019iuo,AraujoFilho:2025rwr,Baruah:2025ifh,AraujoFilho:2025hnf}
\begin{equation}
  V_s(r)=f(r)\left[\frac{\ell(\ell+1)}{r^2}+\frac{1-s^2}{r}f'(r)\right].
  \label{eq:potential}
\end{equation}

{\color{black}For $s=0$ and $s=1$, Eq.~\eqref{eq:potential} represents the usual test scalar and electromagnetic effective potentials on the fixed background.  For $s=2$, the same expression reduces to the Schwarzschild Regge--Wheeler potential in the vacuum limit.  However, in the present spacetime the background is supported by an anisotropic effective matter sector.  A complete gravitational perturbation analysis would therefore require perturbing the metric and the matter variables consistently.  Since this coupled system is not constructed here, the $s=2$ channel used below should be regarded as a phenomenological Regge--Wheeler-type axial channel rather than as the full gravitational perturbation spectrum of the de Sitter-core black hole.}

\begin{figure}[t]
  \centering
  \includegraphics[width=0.92\linewidth]{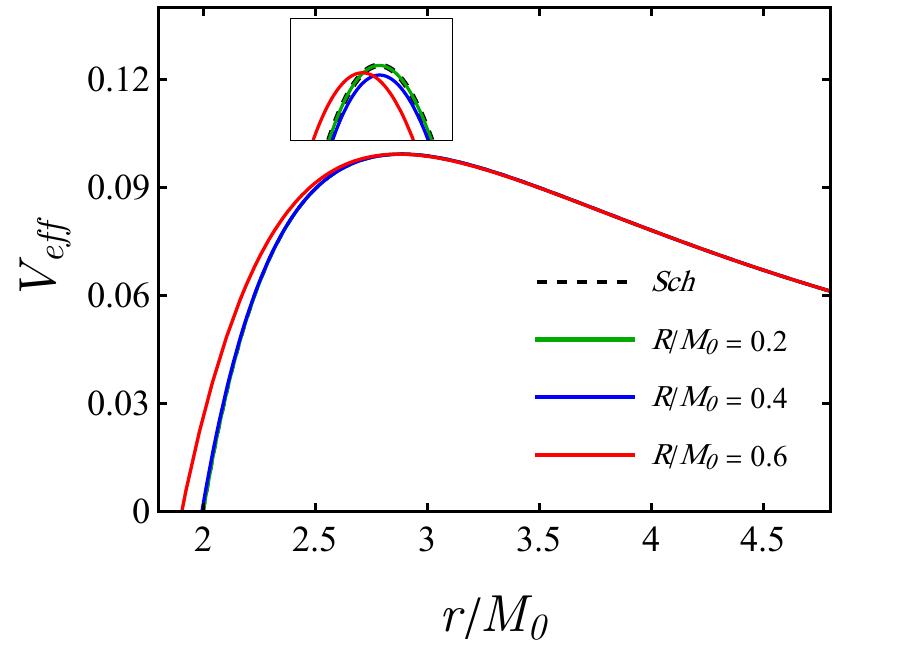}
  \caption{Scalar effective potential for $\ell=1$.  The dashed line denotes the Schwarzschild limit.  Increasing $R/M_0$ modifies the barrier height and position, which in turn shifts the QNM spectrum.}
  \label{fig:Veff}
\end{figure}

\section{Quasinormal modes and greybody factor}
\label{sec:qnm}

The QNMs satisfy ingoing boundary conditions at the event horizon and outgoing boundary conditions at spatial infinity.  We use the third-order WKB formula \cite{iyer1987black,iyer1987black2,konoplya2019higher}
\begin{equation}
  \omega^2=V_0+\sqrt{-2V_0''}\,\Lambda_n-i\alpha\sqrt{-2V_0''}\left(1+\Omega_n\right),
  \label{eq:wkb3}
\end{equation}
where $\alpha=n+1/2$, $n$ is the overtone number, and all derivatives are evaluated at the maximum of the potential with respect to $r_*$.  The correction terms are
\begin{align}
  \Lambda_n=&\frac{1}{\sqrt{-2V_0''}}\left[\frac{1}{8}\frac{V_0^{(4)}}{V_0''}\left(\frac14+\alpha^2\right)
  -\frac{1}{288}\left(\frac{V_0'''}{V_0''}\right)^2\left(7+60\alpha^2\right)\right],
  \label{eq:Lambda}
\end{align}
\begin{align}
  \Omega_n=&\frac{1}{-2V_0''}\Bigg[
  \frac{5}{6912}\left(\frac{V_0'''}{V_0''}\right)^4(77+188\alpha^2)
 \nonumber\\
  &  -\frac{1}{384}\frac{(V_0''')^2V_0^{(4)}}{(V_0'')^3}(51+100\alpha^2)
  \nonumber\\
  &+\frac{1}{2304}\left(\frac{V_0^{(4)}}{V_0''}\right)^2(67+68\alpha^2)
  \nonumber\\
  & +\frac{1}{288}\frac{V_0'''V_0^{(5)}}{(V_0'')^2}(19+28\alpha^2)
  \nonumber\\
  & -\frac{1}{288}\frac{V_0^{(6)}}{V_0''}(5+4\alpha^2)
  \Bigg].
  \label{eq:Omega}
\end{align}
The WKB expansion is most reliable for $\ell>n$ and should be used with caution for the lowest multipoles.

\begin{figure}[t]
  \centering
  \includegraphics[width=0.48\linewidth]{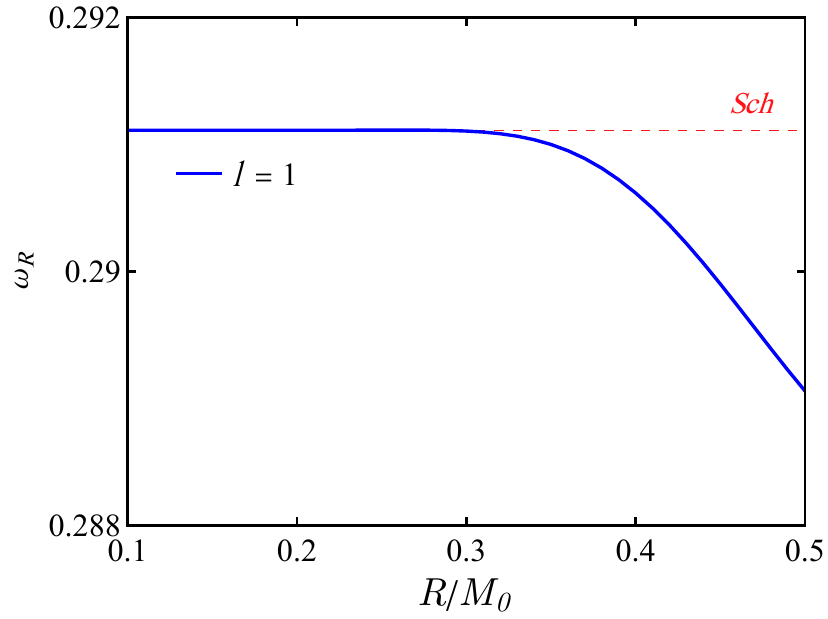}
  \includegraphics[width=0.48\linewidth]{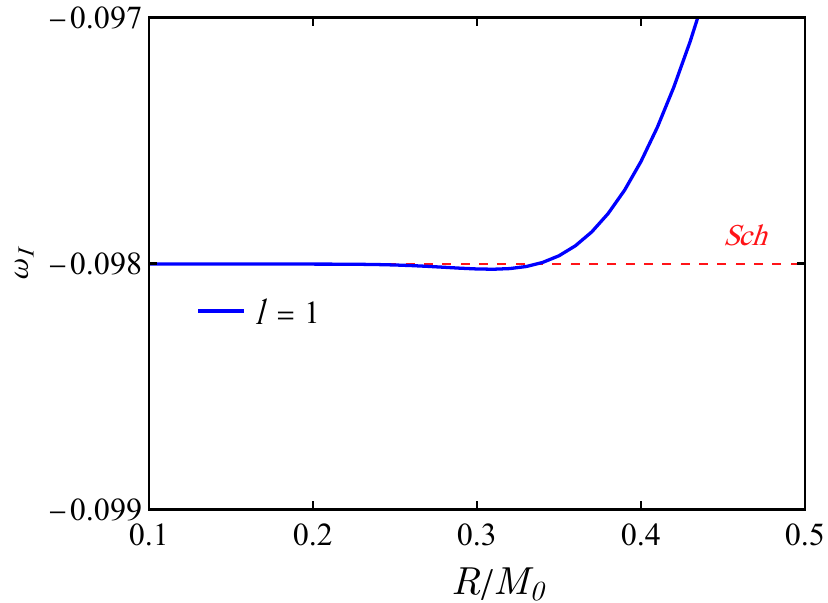}
  \caption{Real part and damping rate of the scalar QNM frequency for $\ell=1$ as functions of $R/M_0$.  The dashed line indicates the Schwarzschild value.  We use the convention $\omega=\omega_R-i\omega_I$, with $\omega_I>0$.}
  \label{fig:QNML1}
\end{figure}

For small $R/M_0$ the effective potential is almost indistinguishable from the Schwarzschild one outside the event horizon, and the fundamental scalar modes remain close to their Schwarzschild values.  As $R/M_0$ approaches the near-extremal regime, the potential barrier is lowered and displaced.  Consequently, both the oscillation frequency $\omega_R$ and the damping rate $\omega_I$ decrease.  The latter implies longer-lived scalar perturbations.

To quantify the deviation from the Schwarzschild spectrum we define
\begin{equation}
  \delta\omega_R=\frac{\omega_R(R)-\omega_R^{\rm Sch}}{\omega_R^{\rm Sch}},
  \qquad
  \delta\omega_I=\frac{\omega_I(R)-\omega_I^{\rm Sch}}{\omega_I^{\rm Sch}}.
  \label{eq:normalized_deviation}
\end{equation}
The fractional shift is largest for lower multipoles, where the wave probes a wider region of the effective potential.  For fixed $\ell$, higher overtones are more sensitive to the deformation, although the WKB accuracy also deteriorates as $n$ approaches $\ell$.

\begin{figure}[t]
  \centering
  \includegraphics[width=0.48\linewidth]{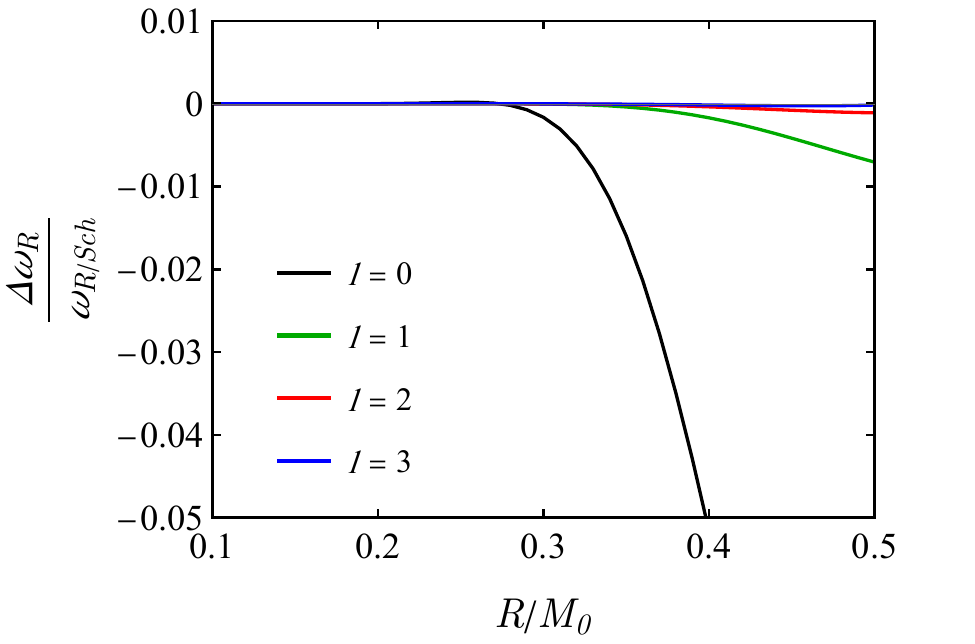}
  \includegraphics[width=0.48\linewidth]{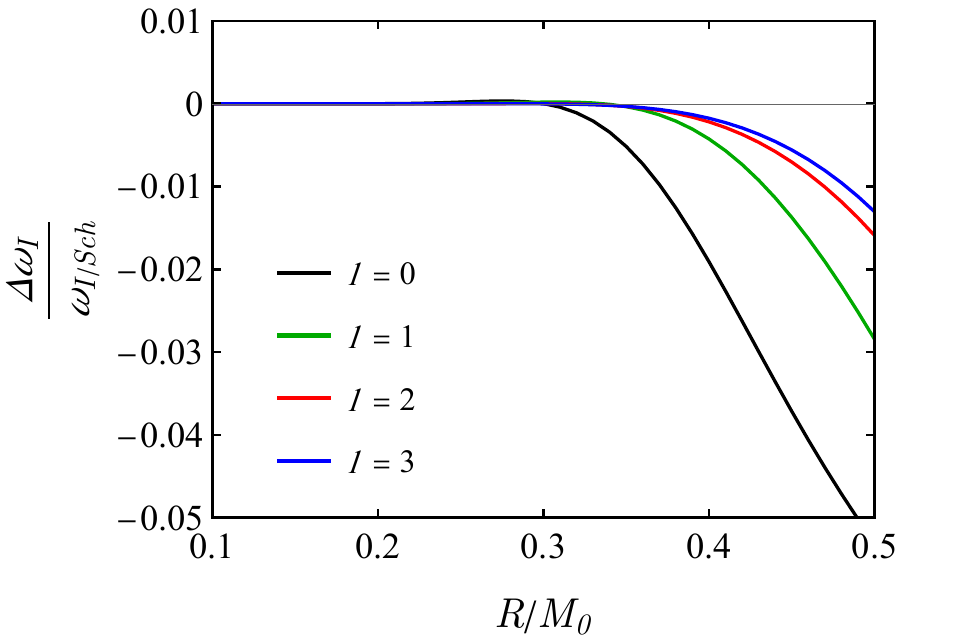}
  \caption{Normalized deviations of the real part and damping rate from the Schwarzschild values for different multipoles.  Curves outside the WKB domain $\ell>n$ should be excluded or interpreted only qualitatively.}
  \label{fig:QNMLIR}
\end{figure}

\begin{figure}[t]
  \centering
  \includegraphics[width=0.48\linewidth]{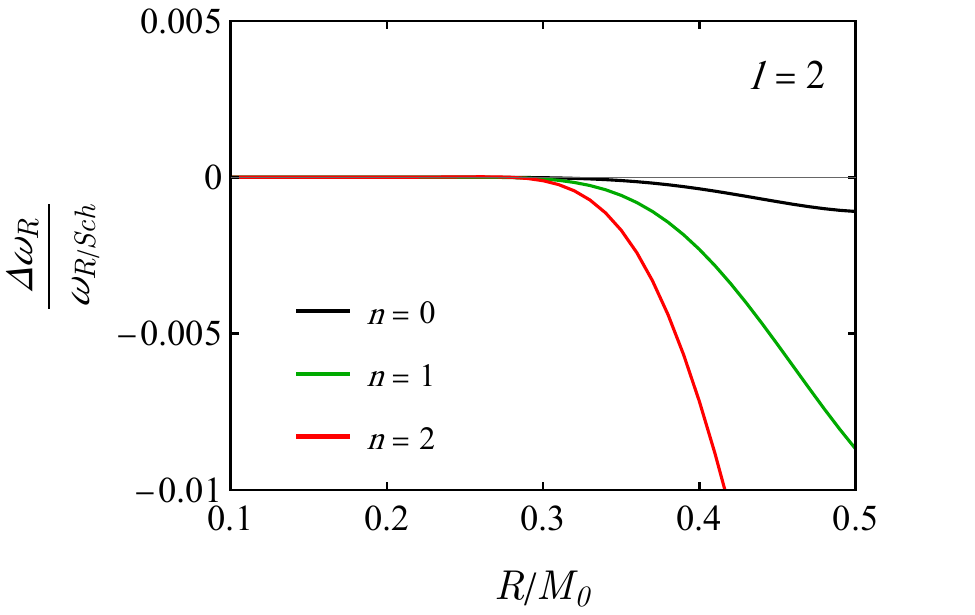}
  \includegraphics[width=0.48\linewidth]{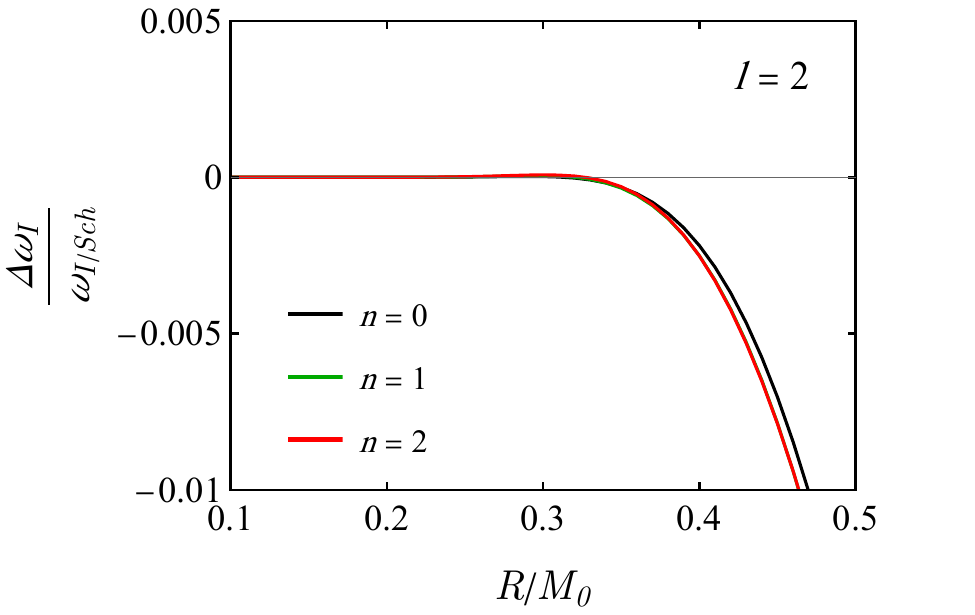}
  \caption{Normalized deviations of the real part and damping rate for $\ell=2$ and $n=0,1$.  The $n=2$ WKB point is not quantitatively reliable and should be used only as a qualitative indication if retained in the figure.}
  \label{fig:QNML2}
\end{figure}

The greybody factor measures the transmission probability through the curvature-induced potential barrier.  We use the rigorous lower-bound method of Refs.~\cite{visser1999some,boonserm2008bounding,Boonserm:2008zg,Ovgun:2023ego},
\begin{equation}
  T_b(\omega)\ge \operatorname{sech}^2\left[\int_{-\infty}^{+\infty}\mathcal{G}(r_*)\mathrm{d}r_*\right],
  \label{eq:gb_bound_general}
\end{equation}
where
\begin{equation}
  \mathcal{G}=\frac{\sqrt{(h')^2+\left(\omega^2-V_s-h^2\right)^2}}{2h}.
  \label{eq:G_function}
\end{equation}
Here $h(r_*)$ is any positive function satisfying $h(\pm\infty)=\omega$.  Choosing $h=\omega$ gives
\begin{equation}
  T_b(\omega)\ge \operatorname{sech}^2\left[\frac{1}{2\omega}\int_{r_+}^{\infty}\left(\frac{\ell(\ell+1)}{r^2}+\frac{1-s^2}{r}f'(r)\right)\mathrm{d}r\right],
  \label{eq:gb_bound_homega}
\end{equation}
provided the effective integrand is non-negative in the parameter range considered.  For scalar perturbations this becomes
\begin{align}
  T_b(\omega)\ge \operatorname{sech}^2\Bigg\{\frac{1}{2\omega}
  \int_{r_+}^{\infty}&\bigg[\frac{\ell(\ell+1)}{r^2}+\frac{2M_0}{r^3}
  \nonumber\\
  &-\frac{2M_0e^{-2r/R}(2r+R)(2r^2+R^2)}{r^3R^3}\bigg]\mathrm{d}r\Bigg\}.
  \label{eq:scalar_gb_bound}
\end{align}

\begin{figure}[t]
  \centering
  \includegraphics[width=0.92\linewidth]{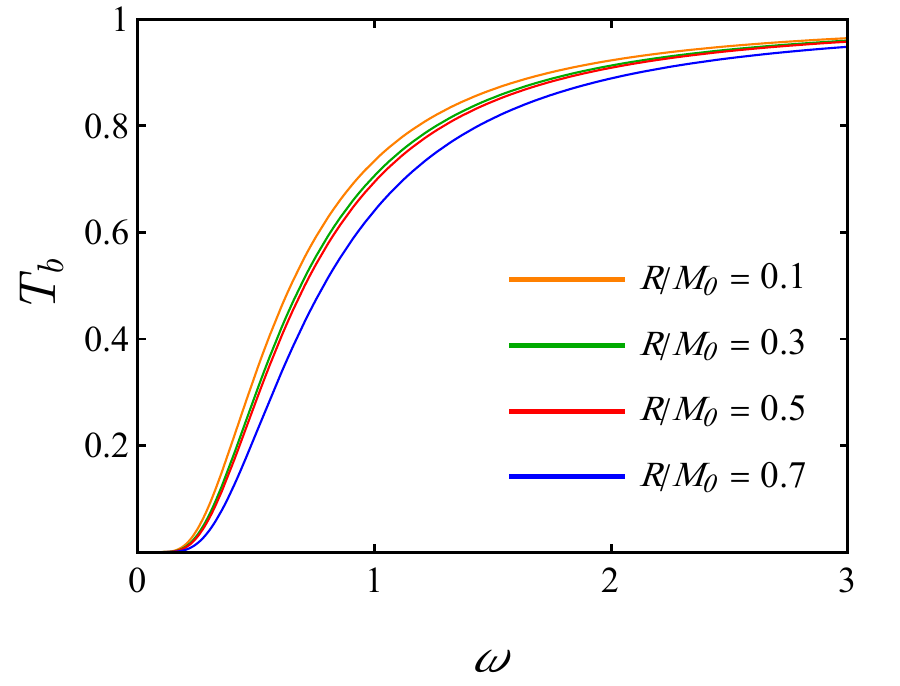}
  \caption{Semi-analytic lower bound for the scalar greybody factor at representative values of $R/M_0$.}
  \label{fig:Tb}
\end{figure}

At fixed frequency, increasing the de Sitter-core scale increases the exponent in Eq.~\eqref{eq:scalar_gb_bound} for the parameter range shown in Fig.~\ref{fig:Tb}.  The lower bound therefore decreases, indicating stronger filtering by the effective potential barrier.  Since Eq.~\eqref{eq:scalar_gb_bound} is a bound rather than an exact transmission coefficient, this conclusion should be interpreted as a conservative statement about the escape probability.

An approximate correspondence between QNMs and greybody factors was recently proposed in Ref.~\cite{konoplya2024correspondence}.  In the WKB language the reflection and transmission coefficients can be written as \cite{iyer1987black,Heidari:2026swh,Heidari:2024bvd}
\begin{equation}
  |R|^2=\frac{1}{1+e^{-2\pi i\mathcal{K}}},
  \qquad
  |T|^2=\frac{1}{1+e^{2\pi i\mathcal{K}}}.
  \label{eq:RT_coefficients}
\end{equation}
Defining $\kappa=-i\mathcal{K}$, the transmission probability is $|T|^2=(1+e^{-2\pi\kappa})^{-1}$.  The correspondence approximates $\kappa$ in terms of the fundamental mode $\omega_0$ and the first overtone $\omega_1$:
\begin{equation}
  \kappa=-\frac{\omega^2-\omega_{0R}^2}{4\omega_{0R}\omega_{0I}}+\Delta_1+\Delta_2+\Delta_f,
  \label{eq:kappa_qnm}
\end{equation}
with
\begin{equation}
  \Delta_1=\frac{\omega_{0R}-\omega_{1R}}{16\omega_{0I}},
  \label{eq:Delta1}
\end{equation}
\begin{align}
  \Delta_2=&-\frac{\omega^2-\omega_{0R}^2}{32\omega_{0R}\omega_{0I}}\left[\frac{(\omega_{0R}-\omega_{1R})^2}{4\omega_{0I}^2}-\frac{3\omega_{0I}-\omega_{1I}}{3\omega_{0I}}\right]
  \nonumber\\
  &+\frac{(\omega^2-\omega_{0R}^2)^2}{16\omega_{0R}^3\omega_{0I}}\left[1+\frac{\omega_{0R}(\omega_{0R}-\omega_{1R})}{4\omega_{0I}^2}\right],
  \label{eq:Delta2}
\end{align}
\begin{align}
  \Delta_f=&-\frac{(\omega^2-\omega_{0R}^2)^3}{32\omega_{0R}^5\omega_{0I}}
  \Bigg[1+\frac{\omega_{0R}(\omega_{0R}-\omega_{1R})}{4\omega_{0I}^2}
  \nonumber\\
  &+\omega_{0R}^2\left(\frac{(\omega_{0R}-\omega_{1R})^2}{16\omega_{0I}^4}-\frac{3\omega_{0I}-\omega_{1I}}{12\omega_{0I}^3}\right)\Bigg].
  \label{eq:Deltaf}
\end{align}
The expression is expected to work best in the eikonal regime; at low $\ell$ it should be viewed as an approximation supplemented by WKB information from the first few overtones.

\begin{figure}[t]
  \centering
  \includegraphics[width=0.92\linewidth]{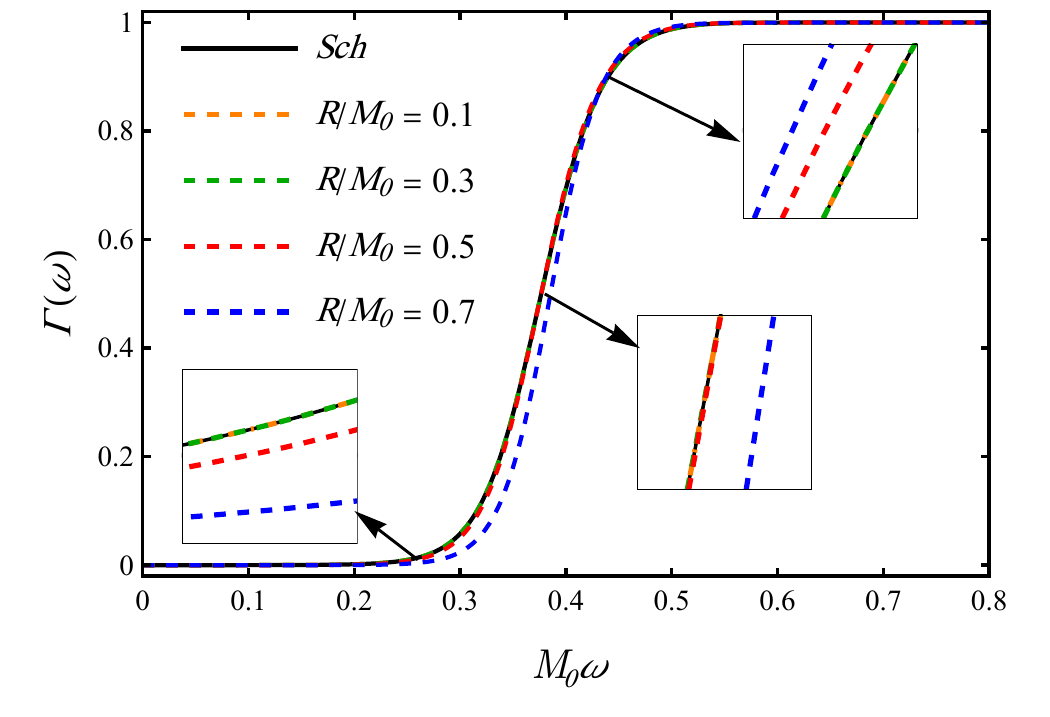}
  \caption{{\color{black}QNM-based estimate of the greybody factor in the phenomenological Regge--Wheeler-type axial channel with $\ell=2$ and representative values of $R/M_0$.}}
  \label{fig:GBFQNM}
\end{figure}

For the de Sitter-core geometry, the QNM-based transmission coefficient is sensitive to the frequency interval.  At frequencies near the peak of the potential barrier, small changes in $R/M_0$ can noticeably shift the transition between reflection-dominated and transmission-dominated propagation.  This behavior is consistent with the deformation of the effective potential shown in Fig.~\ref{fig:Veff}.

The Hawking temperature follows from the surface gravity,
\begin{equation}
  T_H=\frac{f'(r_+)}{4\pi}.
  \label{eq:TH_definition}
\end{equation}
Solving $f(r_+)=0$ for the mass parameter gives
\begin{equation}
  M_0=\frac{R^2r_+e^{2r_+/R}}{2\left(R^2e^{2r_+/R}-2Rr_+-2r_+^2-R^2\right)}.
  \label{eq:M_horizon}
\end{equation}
Substitution into Eq.~\eqref{eq:TH_definition} yields
\begin{equation}
  T_H=\frac{-R^3e^{2r_+/R}+R^3+2R^2r_++2Rr_+^2+4r_+^3}{4\pi Rr_+\left[-R^2e^{2r_+/R}+R^2+2Rr_++2r_+^2\right]}.
  \label{eq:TH_explicit}
\end{equation}
The Schwarzschild result is recovered in the limit $R/M_0\to0$.  At the extremal configuration in Eq.~\eqref{eq:extremal_values}, the temperature vanishes.

\begin{figure}[t]
  \centering
  \includegraphics[width=0.92\linewidth]{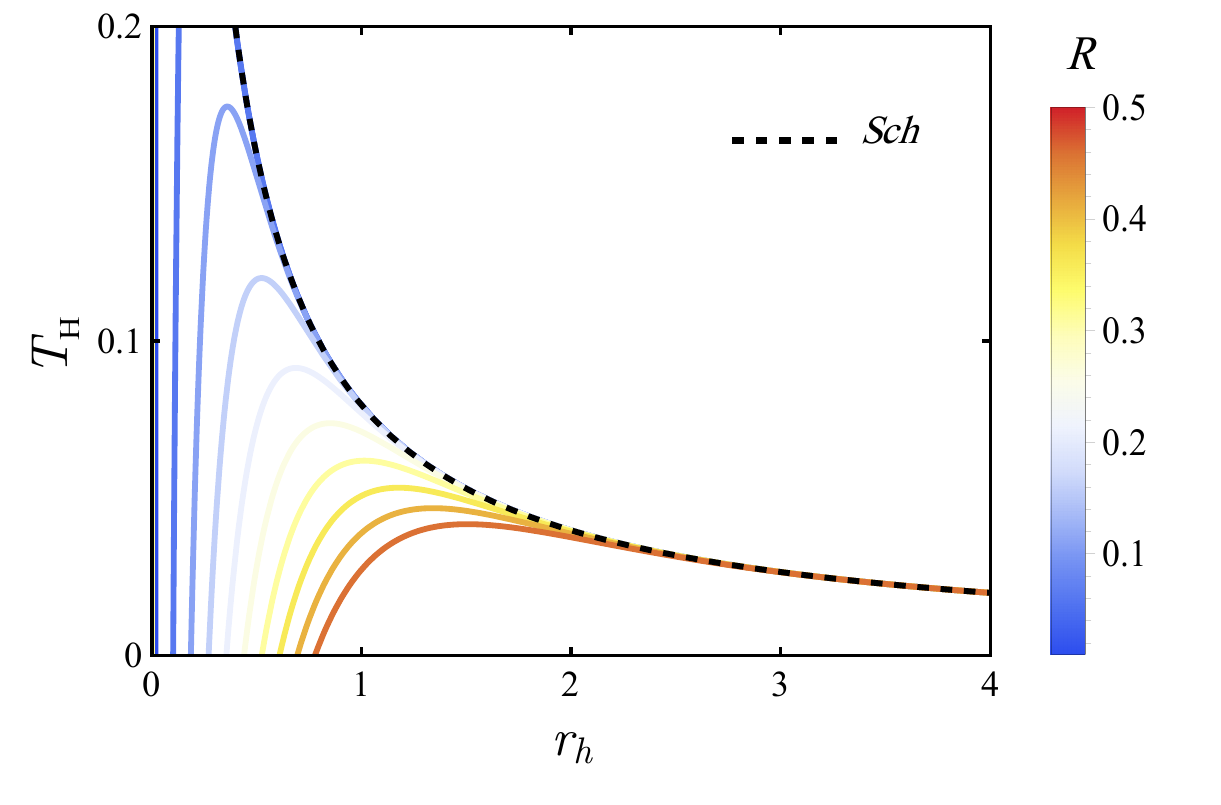}
  \caption{Hawking temperature as a function of the outer horizon radius.  The dashed line denotes the Schwarzschild result.  The temperature tends to zero as the extremal de Sitter-core configuration is approached.}
  \label{fig:Temp}
\end{figure}

The high-frequency absorption cross section is controlled by the critical impact parameter.  For null geodesics in Eq.~\eqref{eq:metric}, the photon-sphere radius satisfies
\begin{equation}
  2f(r_{\rm ph})-r_{\rm ph}f'(r_{\rm ph})=0,
  \label{eq:photon_sphere}
\end{equation}
with shadow radius
\begin{equation}
  b_c=\frac{r_{\rm ph}}{\sqrt{f(r_{\rm ph})}},
  \qquad
  \sigma_{\rm lim}\simeq\pi b_c^2.
  \label{eq:shadow_radius}
\end{equation}
The energy emission rate is then approximated by \cite{decanini2011universality,papnoi2022rotating,AraujoFilho:2025zzf,AraujoFilho:2025hkm,AraujoFilho:2024ctw}
\begin{equation}
  \frac{\mathrm{d}^2E}{\mathrm{d}\omega \mathrm{d}t}=\frac{2\pi^2\sigma_{\rm lim}\omega^3}{e^{\omega/T_H}-1}.
  \label{eq:emission}
\end{equation}

\begin{figure}[t]
  \centering
  \includegraphics[width=0.92\linewidth]{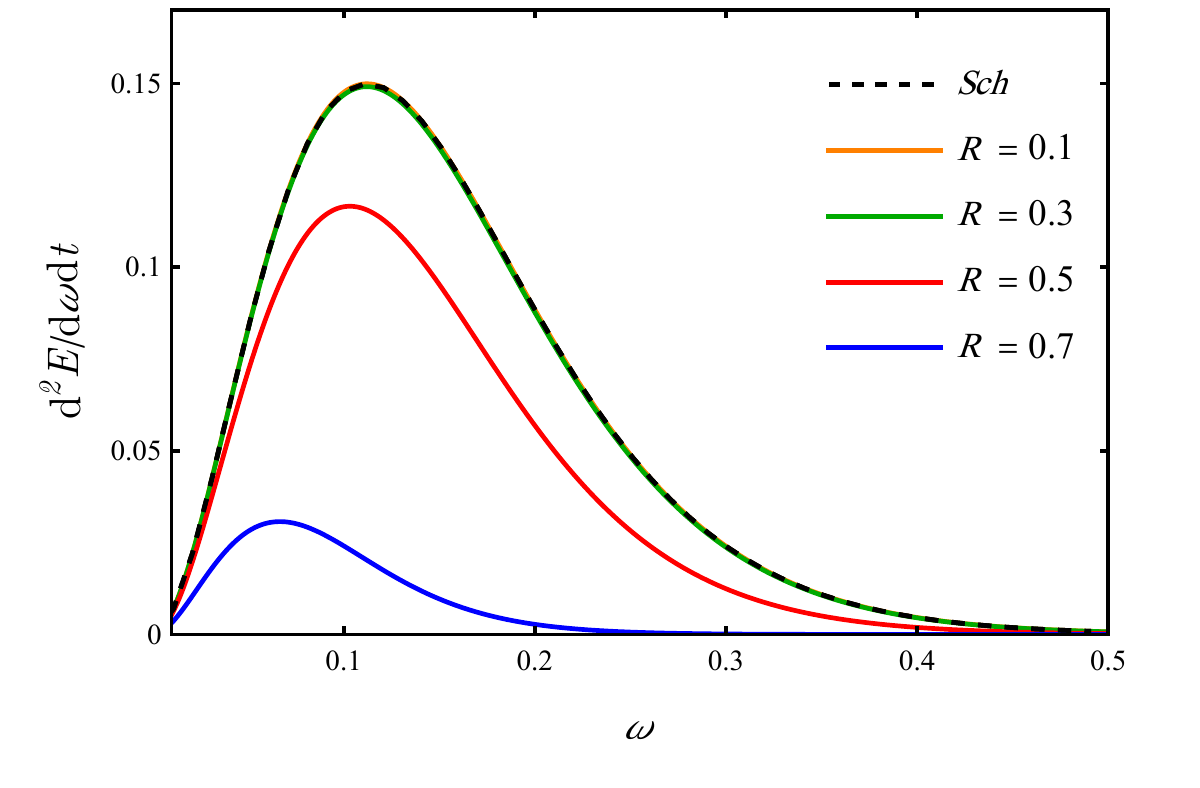}
  \caption{Emission rate for the Schwarzschild black hole and for the de Sitter-core geometry at representative values of $R/M_0$.}
  \label{fig:Emission}
\end{figure}

As $R/M_0$ increases, the Hawking temperature decreases and the emission peak shifts to lower frequencies.  This indicates a slower evaporation rate as the extremal remnant-like configuration is approached.  Other signatures of related regular or dynamical black holes have been explored in Refs.~\cite{misyura2024non,heydarzade2024dynamical,vertogradov2024generalized,vertogradov2022structure,vertogradov2018eternal}.

\section{Optically thin infalling flow and shadow image}
\label{sec:optical}

We finally consider the apparent image produced by an optically thin, radiating flow around the black hole.  The observed specific intensity at the observed frequency $\nu_{\rm obs}$ is
\begin{equation}
  I_{\rm obs}(\nu_{\rm obs},X,Y)=\int_\gamma g^3j(\nu_e)\,\mathrm{d}l_{\rm prop},
  \label{eq:intensity}
\end{equation}
where $g=\nu_{\rm obs}/\nu_e$ is the redshift factor, $j(\nu_e)$ is the rest-frame emissivity and the integral is evaluated along the photon trajectory \cite{Bambi:2012tg,Okyay:2021nnh}.  For radial free fall from rest at infinity in the metric \eqref{eq:metric},
\begin{equation}
  u_e^t=\frac{1}{f(r)},
  \qquad
  u_e^r=-\sqrt{1-f(r)},
  \qquad
  u_e^\theta=u_e^\phi=0.
  \label{eq:emitter_velocity}
\end{equation}
For an equatorial photon with impact parameter $b=L/E$,
\begin{equation}
  k^t=\frac{E}{f},
  \qquad
  k^r=\pm E\sqrt{1-\frac{f b^2}{r^2}}.
  \label{eq:photon_momentum}
\end{equation}
The corresponding redshift factor can be written as
\begin{equation}
  g=\frac{f(r)}{1\pm\sqrt{1-f(r)}\sqrt{1-f(r)b^2/r^2}},
  \label{eq:redshift_factor}
\end{equation}
where the sign is fixed by the radial direction of the photon.  We use a monochromatic emissivity with radial profile
\begin{equation}
  j(\nu_e)\propto\frac{\delta(\nu_e-\nu_\star)}{r^2}.
  \label{eq:emissivity}
\end{equation}
After integrating over frequency, the observed flux is proportional to
\begin{equation}
  F_{\rm obs}(X,Y)\propto-\int_\gamma\frac{g^3k_t}{r^2k_r}\,\mathrm{d}r.
  \label{eq:flux}
\end{equation}
This expression is evaluated numerically by backward ray tracing.

\begin{figure}[t]
  \centering
  \includegraphics[width=0.92\linewidth]{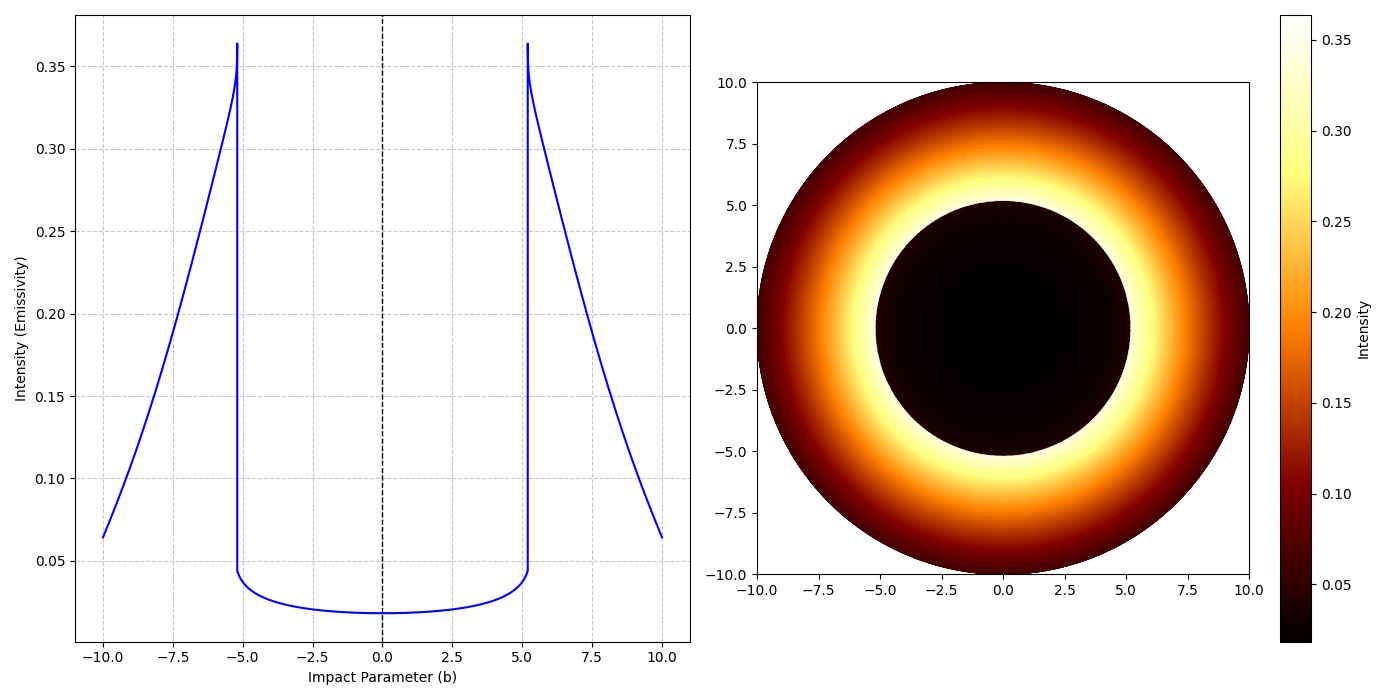}
  \includegraphics[width=0.92\linewidth]{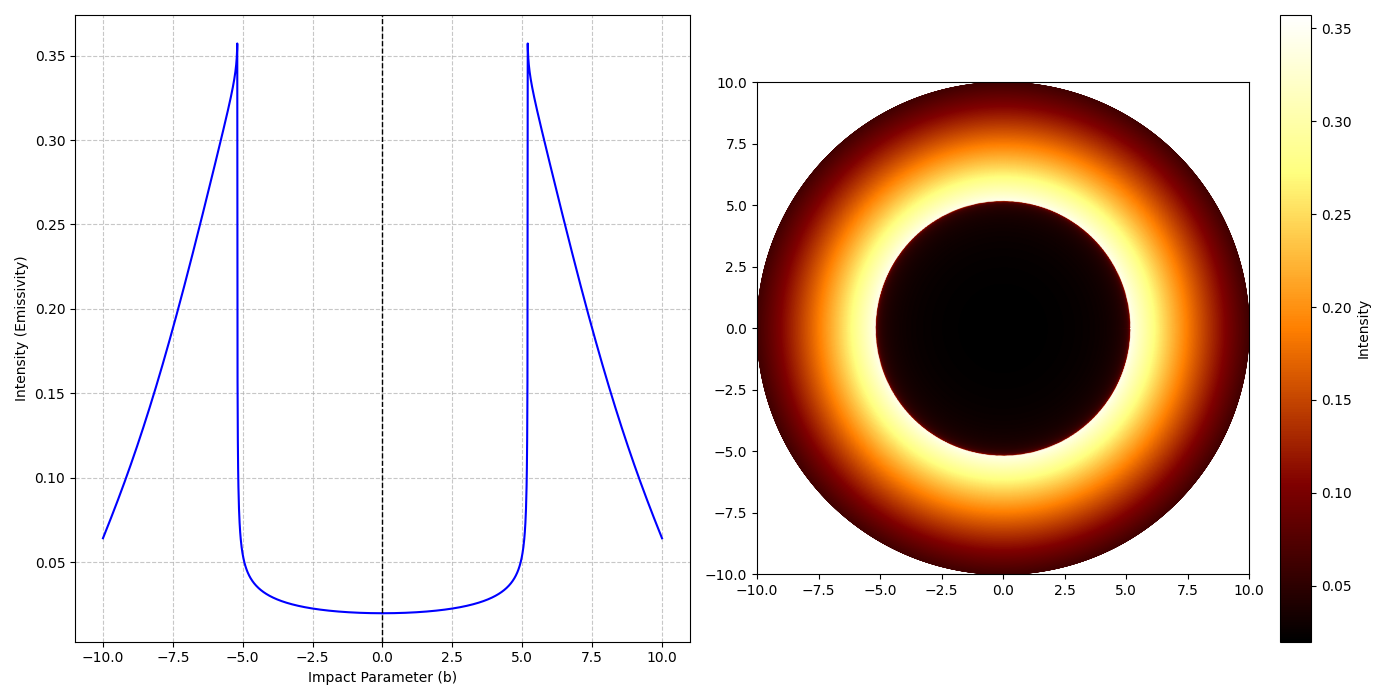}
  \includegraphics[width=0.92\linewidth]{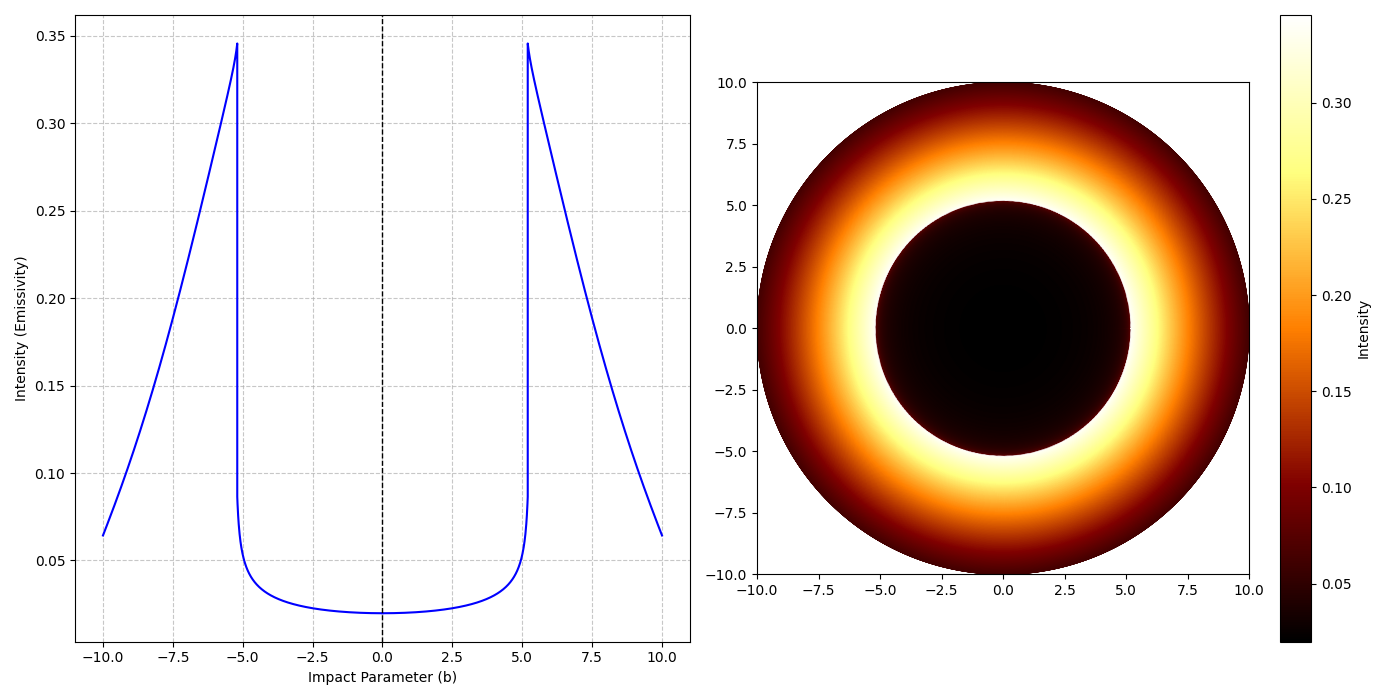}
  \caption{Intensity profiles and shadow images of the de Sitter-core black hole for $M_0=1$ and $R=0.2,0.5,0.7$.  A larger core scale mildly reduces the apparent shadow size and the peak intensity of the infalling-flow image.}
  \label{fig:OpticalImages}
\end{figure}

The images show the expected central brightness depression associated with photon capture.  As $R$ increases, the photon-sphere structure and redshift factor are modified, producing a small reduction in both the apparent shadow scale and the peak observed intensity.

{\color{black}The image model used here in Fig.~\ref{fig:OpticalImages} should be viewed as a controlled geometric test.  It neglects disk thickness, magnetic fields, absorption, electron-temperature structure and instrumental response.  These effects can change the brightness distribution and may partly hide the small geometric shift caused by $R/M_0$. Consequently, observational distinguishability of the de Sitter-core parameter requires dedicated radiative-transfer modelling and should be regarded as model-dependent.}

{\color{black}\subsection{Realistic thin-disk model}
\label{ssec:GLM}

The radial in-fall analysis of the previous paragraph is a clean
\emph{geometric} diagnostic, but it lacks the photon-ring brightness
enhancement that dominates real horizon-scale images.  We therefore
also examine the optical appearance produced by an optically thin,
geometrically thin equatorial disk \cite{Gralla:2020srx,Vincent:2022fwj}.
For an impact parameter $b$, the observed intensity reads
\begin{equation}
  I_{\rm obs}(b) \;=\; \sum_{m=0}^{2}\,\bigl[g(r_m)\bigr]^{3}\,j(r_m),
  \qquad
  g(r)=\sqrt{f(r)},
  \label{eq:I_GLM}
\end{equation}
where $r_m(b)$ is the radius at which the photon crosses the equatorial
plane for the $m$-th time, computed from the Binet equation
$\mathrm{d}^{2}u/\mathrm{d}\phi^{2}=-u f(1/u)+\tfrac{1}{2}f'(1/u)$
with $u=1/r$.  The $m=0,1,2$ branches correspond to the direct image,
the lensing ring, and the photon ring, respectively.

To bracket the astrophysical systematics we adopt two
representative emission profiles of Johnson--SU type,
\begin{equation}
  j(r)=\frac{\exp\!\bigl\{-\tfrac{1}{2}[\gamma+\sinh^{-1}\!((r-\mu)/\sigma)]^{2}\bigr\}}
            {\sqrt{(r-\mu)^{2}+\sigma^{2}}},
  \label{eq:jsu}
\end{equation}
namely Profile~1 with $(\mu,\gamma,\sigma)=(r_{\rm ISCO},-2,0.5\,M_0)$ ---
a sharp peak just outside the innermost stable circular orbit
appropriate for a truncated, radiatively efficient disk --- and
Profile~2 with $(\mu,\gamma,\sigma)=(\tfrac{1}{2}(r_{+}+r_{\rm ISCO}),0,M_0)$ ---
an emission concentrated between the horizon and the ISCO and peaking
close to the photon sphere.  The ISCO is obtained from
$r f f''+3ff'-2r(f')^{2}=0$ and is nearly insensitive to $R/M_{0}$:
we find $r_{\rm ISCO}/M_{0}=6.000,\;5.99998,\;5.994$ for
$R/M_{0}=0.2,0.5,0.7$ respectively.

\begin{figure*}[t]
  \centering
  \includegraphics[width=0.98\textwidth]{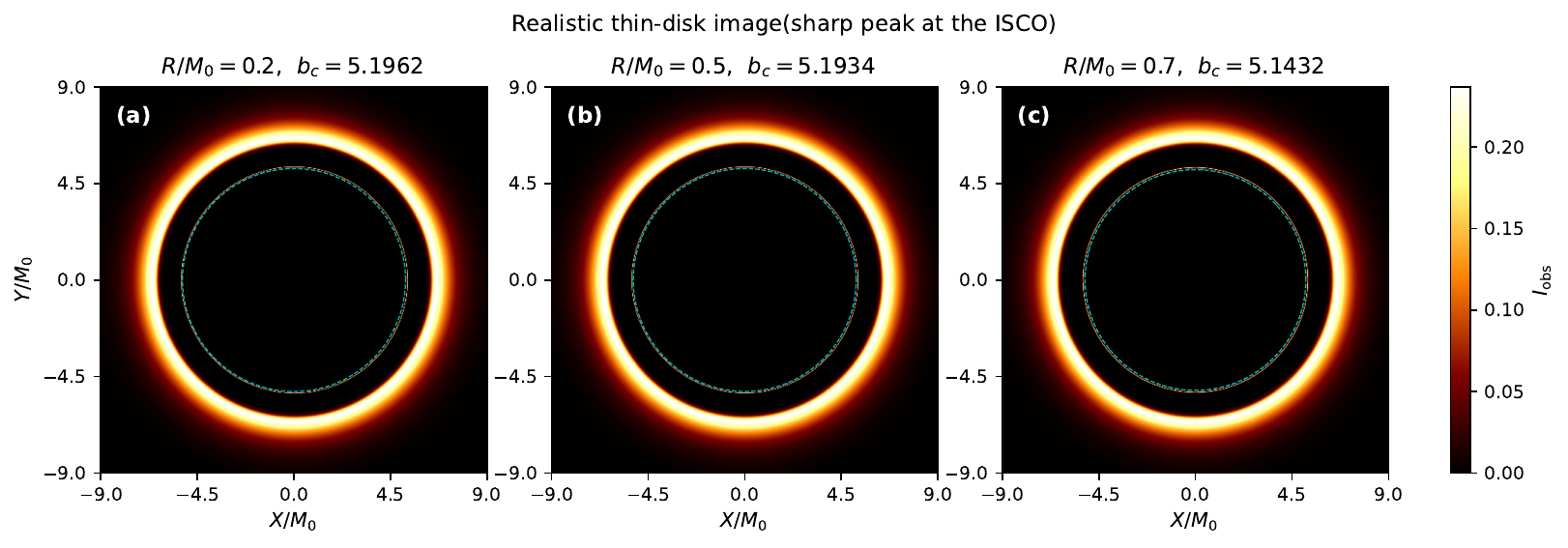}
  \caption{Realistic GLM thin-disk image of the de Sitter-core black
    hole for Emission Profile~1, in which the rest-frame emissivity
    peaks sharply just outside the ISCO.  Panels (a)--(c) correspond
    to $R/M_{0}=0.2,\,0.5,\,0.7$.  The bright annulus around the
    central depression is the unresolved superposition of the direct
    image, the lensing ring, and the photon ring; the cyan dashed
    circle marks the critical impact parameter $b_{c}$.  The
    photon-ring scale contracts mildly as $R/M_{0}$ grows.}
  \label{fig:Profile1}
\end{figure*}

\begin{figure*}[t]
  \centering
  \includegraphics[width=0.98\textwidth]{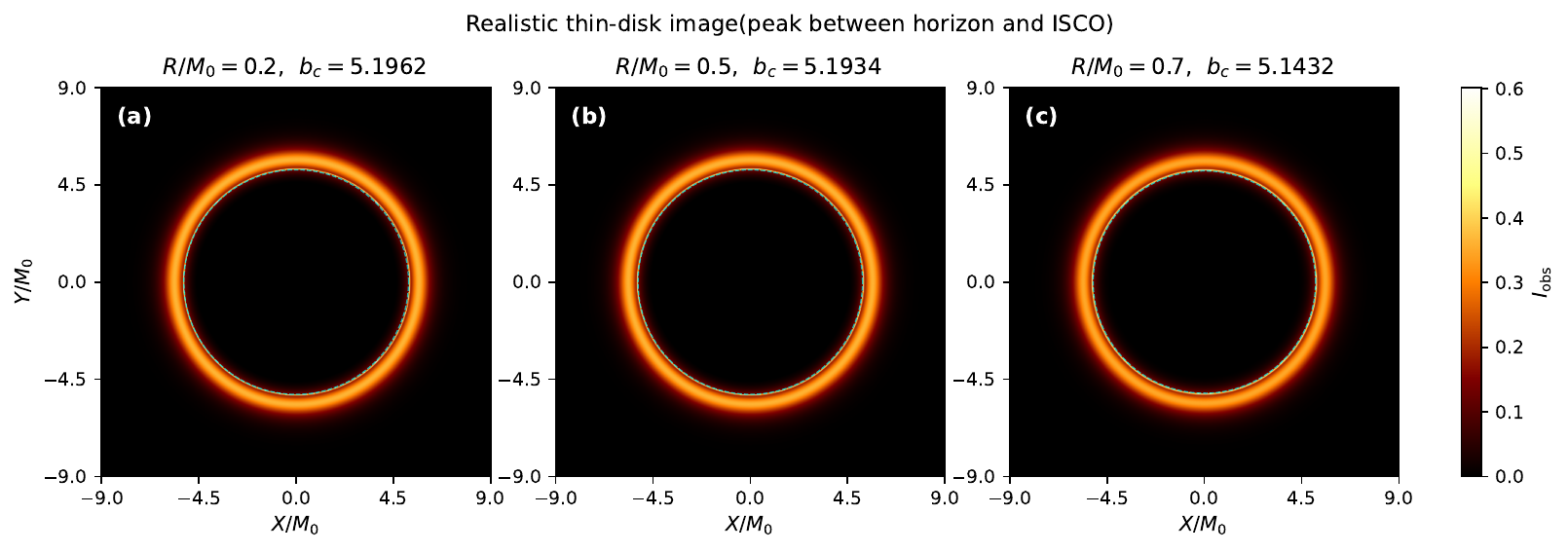}
  \caption{Same as Fig.~\ref{fig:Profile1} but for Emission Profile~2,
    in which the rest-frame emissivity peaks between the event horizon
    and the ISCO.  The image is dominated by emission lensed close to
    the photon sphere, producing a narrower and sharper ring at
    $b\simeq b_{c}$ with a fainter outer halo than in
    Fig.~\ref{fig:Profile1}.  The shadow scale contracts in the same
    sense as for Profile~1 as $R/M_{0}$ grows.}
  \label{fig:Profile2}
\end{figure*}

Figures~\ref{fig:Profile1} and~\ref{fig:Profile2} show the
backward-ray-traced images obtained from
Eqs.~\eqref{eq:I_GLM}--\eqref{eq:jsu} for both profiles at
$R/M_{0}=0.2,\,0.5,\,0.7$.  Profile~1 produces a bright, thick annulus
because the emissivity peaks near the inner edge of an extended disk:
photons reach the observer through the direct image, the lensing ring
and the photon ring, and the three contributions partly overlap.
Profile~2 instead concentrates the emission close to the photon
sphere, so the direct contribution is dim and the observed image is
strongly dominated by the lensed component, yielding a thinner and
sharper ring at $b\simeq b_{c}$.  In both cases the apparent shadow
scale set by $b_{c}$ decreases mildly as $R/M_{0}$ grows, consistent
with the simple-model trend of Fig.~\ref{fig:OpticalImages}.

Quantitatively, the critical impact parameter shifts from
$b_{c}/M_{0}=5.1962$ at $R/M_{0}=0.2$ to $5.1934$ at $R/M_{0}=0.5$ and
$5.1432$ at $R/M_{0}=0.7$, i.e.\ a fractional deviation
$|\Delta b_{c}/b_{c}^{\rm Sch}|$ of $5\times10^{-4}$ and $1.0\%$
respectively, reaching at most $\sim2.1\%$ at the extremal value
$R/M_{0}=R_{e}/M_{0}\simeq0.7768$.  Such a shift is smaller than the
$1\sigma$ uncertainty on the M87$^{*}$ emission-ring diameter measured
by the Event Horizon Telescope, which is approximately
$6.7\%$~\cite{EventHorizonTelescope:2019dse}, and is also below the
$\sim10\%$ combined uncertainty for Sgr~A$^{*}$~\cite{EventHorizonTelescope:2022wkp}.
The de Sitter-core deformation is therefore geometrically
degenerate with Schwarzschild within current horizon-scale imaging
precision: identifying a non-zero $R/M_{0}$ from the shadow alone
would require sub-percent measurements of the photon-ring diameter,
attainable in principle by future space-VLBI missions, or a joint
analysis combining horizon-scale imaging with the ringdown signal of
Sec.~\ref{sec:qnm}, where the deviations are several times larger
than in the imaging channel.
}


\section{Conclusions}
\label{sec:conclusions}

We have studied a static, asymptotically flat black hole with a de Sitter core and examined its implications for perturbations, transmission, evaporation and optical appearance.  The solution interpolates between a de Sitter-like center and a Schwarzschild exterior.  Its effective source is anisotropic, with $p_r=-\rho$ and $p_\perp=(r/R-1)\rho$, and the transverse equation of state can be written covariantly as Eq.~\eqref{eq:eos_covariant}.  The horizon analysis shows that two horizons exist for $R/M_0\lesssim0.7768$ and merge at an extremal configuration where the Hawking temperature vanishes.

For scalar perturbations, the de Sitter core modifies the effective potential mainly when $R/M_0$ becomes appreciable.  The WKB analysis indicates that the real oscillation frequency and damping rate decrease as the extremal regime is approached.  The fractional shift is largest for low multipoles and for higher overtones within the range where the WKB approximation remains trustworthy.  The greybody-factor bound shows that a larger core scale increases the filtering effect of the potential barrier, while the QNM--greybody correspondence provides a complementary estimate of the transmission probability in the eikonal regime.

Thermodynamically, the de Sitter core suppresses the Hawking temperature and shifts the emission maximum toward lower frequencies, suggesting slower evaporation near the extremal endpoint.  The optical analysis of an optically thin infalling flow shows a mild decrease in the apparent shadow size and peak intensity as $R$ increases.  These results suggest that ringdown, greybody transmission and horizon-scale imaging provide complementary probes of regular-core black-hole geometries.

{\color{black}The optical calculation gives a modest reduction of the shadow scale and peak intensity as $R/M_0$ increases.  Because the effect is small, imaging alone is unlikely to isolate the core parameter with present accuracy.  A more robust test would combine photon-ring information with the ringdown spectrum, and would require a full perturbation treatment including the anisotropic matter sector.}

\section*{Data availability}
No new observational data were generated or analyzed in this work.  The numerical data used to produce the figures can be regenerated from the equations given in the text.

\section*{Declaration of generative AI use}
During the preparation of this work the author(s) used ChatGTP in order to improve grammar of paper, and assist with the formatting of LATEX code. After using this tool/service, the author(s) reviewed and edited the content as needed and take(s) full responsibility for the content of the published article.  All intellectual content, analysis, and conclusions are the authors’ own.

\section*{Acknowledgments}
A. A. Ara\'{u}jo Filho is supported by Conselho Nacional de Desenvolvimento Cient\'{i}fico e Tecnol\'{o}gico (CNPq) -- [150223/2025-0].  N. H. acknowledges the COST Action CA21106, COSMIC WISPers in the Dark Universe: Theory, astrophysics and experiments (CosmicWISPers), the COST Action CA21136, Addressing observational tensions in cosmology with systematics and fundamental physics (CosmoVerse), and the COST Action CA23130, Bridging high and low energies in search of quantum gravity (BridgeQG).  This work is supported by Conselho Nacional de Desenvolvimento Cient\'{i}fico e Tecnol\'{o}gico -- CNPq, project number 152891/2025-0.

\bibliographystyle{elsarticle-num}
\bibliography{main}

\end{document}